\documentstyle[12pt]{article}
\begin{document}

\begin{center}
{\large {\bf VOLUME\ INTEGRAL\ THEOREM FOR EXOTIC MATTER}}\\[0pt]

\vspace{8mm}

Kamal Kanti Nandi,$^{a,c,}$\footnote{%
E-mail: kamalnandi@hotmail.com} Yuan-Zhong Zhang$^{b,c,}$\footnote{%
E-mail: yzhang@itp.ac.cn}\\[0pt]
and K.B.Vijaya Kumar$^{d,}$\footnote{%
E-mail: kbvijayakumar@yahoo.com}

\vspace{5mm} {\footnotesize {\it \ $^a$ Department of Mathematics,
University of North Bengal, Darjeeling (W.B.) 734 430, India\\[0pt]
$^b$ CCAST (World Laboratory), P.O.Box 8730, Beijing 100080, China\\[0pt]
$^c$ Institute of Theoretical Physics, Chinese Academy of Sciences, P.O.Box
2735, Beijing 100080, China\\[0pt]
$^d$ Department of Physics, University of Mangalore, Mangalore 574 199, India%
}}
\end{center}

\bigskip

\begin{abstract}
We answer an important question in general relativity about the
volume integral theorem for exotic matter by suggesting an exact
integral quantifier for matter violating Averaged Null Energy
Condition (ANEC). It is checked against some well known static,
spherically symmetric traversable wormhole solutions of general
relativity with a sign reversed kinetic term minimally coupled
scalar field. The improved quantifier is consistent with the
principle that traversable wormholes can be supported by
arbitrarily small quantities of exotic matter.

\bigskip

\noindent PACS number(s): 04.70.Dy,04.62.+v,11.10.Kk
\end{abstract}

\vspace{8mm}

\vspace{20mm} Traversable wormholes are just as good a prediction of
Einstein's general relativity as are black holes. The notion of exotic
matter required to construct such wormholes has found a novel justification
in the role of dark matter on a cosmological scale [1]. There arises a
natural enquiry as to how much of such exotic matter, violating specifically
the Averaged Null Energy Condition (ANEC), is required to support a
traversable Lorentzian wormhole on a local scale. This question has been
recently addressed by Visser, Kar and Dadhich [2]. Their key result is the
proposal of quantifying the total ANEC violating matter in terms of a volume
integral theorem. Such a theorem is of paramount importance as it has
potential implications for topological censorship or positive mass theorems
of general relativity. Moreover, given the widespread interest in
traversable wormholes in recent years, especially among the physics
community, it is imperative that the volume integral in question be properly
identified. The need for a correct quantifier has also been recognized in
Ref. [3].

The work in Ref.[2] concluded with a {\it qualitative} statement about the
total amount of exotic matter which states that appropriately chosen
traversable wormholes can be supported by arbitrarily small quantities of
exotic matter. Let us make it very clear at the outset that we do not
contend this important conclusion. Our interest here is different, that is,
to know the {\it exact quantity }of exotic matter present in a given
spacetime and we feel that a more reasonable approach could be to first
identify the corresponding volume integral and then draw qualitative
conclusions (i.e., large or small quantity) as corollaries. The main purpose
of this Brief Report is an attempt to do just this. We suggest an improved
volume quantifier that checks well against some known examples of static,
spherically symmetric traversable wormholes in the Einstein minimally
coupled scalar field theory. That is to say, the integral satisfies an
important physical criterion in reproducing, to first order, the scalar
\textquotedblleft charge\textquotedblright\ present in the solution, and
hence the exact quantity of exotic matter.

Let us begin with the volume integration measure $dV$, viz.,
\textquotedblleft $4\pi R^{2}dR$". It is a four dimensional natural measure
in the Reissner-Nordstr\"{o}m type of solution in curvature coordinates
where, strictly, $g_{tt}g_{RR}=-1$. But, if the same measure
\textquotedblleft $4\pi R^{2}dR$" is applied to different spherically
symmetric solutions (where $g_{tt}g_{RR}\neq -1$), like the ones we are
going to consider, the integral $4\pi \int {\rho }R^{2}dR$ does not
reproduce the exotic mass (scalar charge in our case). Of course, it may
still approximate the exact value in some way, but, as mentioned, our
interest here is in exact values. Similar comments apply also to the measure
\textquotedblleft $\sqrt{g_{3}}d^{3}x$" (where $g_{3}$ is the determinant of
the spatial part of the metric). All these will be evident in what follows.

Consider the Morris-Thorne-Yurtsever (MTY) [4] form of a static, spherically
symmetric wormhole in the curvature coordinates ($t,R,\theta ,\varphi $) (We
take $G=c=1$):
$$
ds^{2}=-\exp [2\phi ]dt^{2}+\frac{{dR^{2}}}{{1-b(R)/R}}+R^{2}\left( {d\theta
^{2}+\sin ^{2}\theta d\varphi ^{2}}\right) .\eqno(1)
$$%
The throat of the wormhole occurs at $R=R_{0}$ such that $b(R_{0})=R_{0}$,
and we assume ${\rm exp}[2\phi (R_{0})]\neq 0$. The density and pressures
can be calculated in the static orthonormal frame. Then, using these, one
has
$$
\Omega =4\pi \times \int_{R_{0}}^{\infty }{\left[ {\rho +p_{R}}\right] }%
\times R^{2}dR\equiv \left[ {\left( {R-b}\right) \ln \left( {\frac{{\exp
[2\phi ]}}{{1-b/R}}}\right) }\right] _{R_{0}}^{\infty
}-\int\limits_{R_{0}}^{\infty }{\left( {1-b^{\prime }}\right) }\left[ {\ln
\left( {\frac{{\exp [2\phi ]}}{{1-b/R}}}\right) }\right] dR,\eqno(2)
$$%
where the prime denotes differentiation with respect to $R$. Additionally,
the boundary term has been assumed to be zero and the second (integral) part
has been proposed in Ref.[2] as the volume-integral theorem that provides
information about the total amount of ANEC-violating matter in the
spacetime. The boundary term can be made to vanish in several ways. One
possibility is that, $\phi (R)$ and $b(R)$ be asymptotically Schwarzschild
[2], i.e., $\phi (R)\sim -m/R+O(R^{-2})$ and $b(R)\sim 2m+O(R^{-1})$. An
useful alternative could be $\phi (R)\sim O(R^{-2})$ and $b(R)\sim O(R^{-1})$%
, given the possibility of a host of traversable wormholes one is free to
construct. However, for our purposes, we require that the spacetime be only
asymptotically flat although, just incidentally, the examples we are going
to consider follow the first set of asymptotic Schwarzschild behavior.

To get an idea of the value of the charge we want to retrieve, consider the
Einstein minimally coupled scalar field theory given by the field equations
$$
R_{\mu \nu }=-2\Phi _{,\mu }\Phi _{,\nu }\eqno(3)
$$%
$$
\Phi _{;\mu }^{;\mu }=0,\eqno(4)
$$%
where $\mu ,\nu =0,1,2,3$; $\Phi $ is the scalar field, $R_{\mu \nu }$ is
the Ricci tensor and the semicolon denotes covariant derivatives with
respect to the metric $g_{\mu \nu }$. The minus sign on the right implies
that the scalar field has a negative kinetic energy so that the stresses
violate energy conditions (ghost scalar field) [5]. These field equations
are just the vacuum Jordan frame Brans-Dicke equations rewritten in the
conformally rescaled Einstein frame. (They also follow from the vacuum low
energy string theory in four dimensions). Scheel, Shapiro and Teukolsky [6]
have shown that the general asymptotically flat, static solution has the
asymptotic form
$$
g_{00}=-1+\frac{{2M_{T}}}{r},\quad \;g_{0i}=0,g_{ij}=1+\delta _{ij}\left( {%
\frac{{2M_{T}}}{r}}\right) ,\;\quad \Phi =1+\frac{{2M_{S}}}{r},\eqno(5)
$$%
where $i,j=1,2,3$; $r$ is the isotropic radial variable, $M_{T}$ and $M_{S}$
are the tensor and (exotic) scalar masses respectively. $M_{S}$ may be
termed as scalar \textquotedblleft charge" on one side of the wormhole. Our
viewpoint is that it is this $M_{S}$ that the desired quantifier should
first reproduce, and thereby justify itself, before it can be employed to
assess the total ANEC violating matter. To this end, we now state our ANEC
volume integral

$$
\Omega _{ANEC}=\int_{x_{th}}^{\infty }\int \int [{T}_{\mu \nu }k^{\mu
}k^{\nu }{]\sqrt{-g_{4}}d}^{3}x,\eqno(6)
$$%
for null $k^{\mu }$, stress tensor ${T}_{\mu \nu }$, $g_{4}\equiv \det
\left\vert {g_{\mu \nu }}\right\vert $ and the throat at $x_{th}$. Some
additional comments are in order here. We have picked up the integration
measure \textquotedblleft ${\sqrt{-g_{4}}d}^{3}x\textquotedblright $ from
the general relativity conservation law with the difference that the
integration is taken from $x_{th}$ to $\infty $ because of the allowed
coordinate range in wormhole geometry. It should be applicable to any
spacetime that is asymptotically Minkowskian. In the simple case of
spherical symmetry, assuming that the ANEC violating matter is related only
to ${p_{r}}$, and not to the transverse components [2], we have

$$
\Omega _{ANEC}=\int_{r_{0}}^{\infty }\int_{0}^{\pi }\int_{0}^{2\pi }[{\rho
+p_{r}]\sqrt{-g_{4}}drd\theta d\varphi },\eqno(7)
$$%
where $r_{0}$ is the throat radius. We want to try Eq.(7) with some examples
below.

The form of a certain exact general class of solutions of the Eqs. (3) and
(4) is given in isotropic coordinates ($t,r,\theta ,\varphi $) by:
\[
ds^{2}=g_{\mu \nu }dx^{\mu }dx^{\nu }=-e^{2\phi (r)}dt^{2}+e^{-2\psi (r)}
\left[ {dr^{2}+r^{2}d\theta ^{2}+r^{2}\sin ^{2}\theta d\varphi ^{2}}\right]
,
\]%
$$
\phi (r)=\psi (r)=-\frac{M}{r},\quad \quad \Phi (r)=1-\frac{M}{r}.\eqno(8)
$$%
This solution was actually proposed by Yilmaz [7] decades ago, in fact a few
years earlier than the advent of Brans-Dicke theory. However, it follows
also from the Brans-Dicke theory under conformal rescaling. For this
solution $M_{T}=M$ and $M_{S}=-M/2$. The metric in Eqs.(8) exactly coincides
up to second order with the Schwarzschild metric in isotropic coordinates.
That is, the solution describes all the weak field tests of general
relativity just as exactly as the Schwarzschild metric does for $r>M/2$. It
is actually a singularity free solution as the curvature scalars are all
zero at $r=0$ and at $r=\infty $, and thus the solution has two
asymptotically flat regions. The tidal forces are finite everywhere. In
fact, it satisfies all the five conditions laid down by Visser [8] for any
isotropic form to qualify as a traversable wormhole (see Ref. [9] for more
details). The throat appears at $r_{0}=M$. Calculations of the energy
density ($\rho $) and pressures ($p_{r},p_{\theta },p_{\varphi }$) give $%
\rho =-f,p_{r}=-f,p_{\theta }=p_{\varphi }=f$ where $f\equiv (\frac{1}{8\pi }%
)M^{2}r^{-4}e^{-2M/r}>0$. That is, both the Weak Energy Condition ($\rho
\geq 0$) and NEC (${\rho +p_{r}\geq 0}$) are violated, as expected in a
spacetime containing wormholes. With these expressions, the integral (7)
converges and immediately gives the values for the scalar charge $\Omega
_{ANEC}^{p_{r}=0}=-M/2\equiv M_{S}$, and the total ANEC violating mass $%
\Omega _{ANEC}=-M$, no matter whatever coordinate network we use. These
results fundamentally confirm the validity of our integral. Returning to the
MTY form via the transformation $R=r{\rm exp}[M/r]$ (Note that for both $%
r\rightarrow 0$ and $r\rightarrow \infty $, we have $R\rightarrow \infty $
and the throat now occurs at $R_{0}=Me$), and calculating with (2), we find $%
\Omega ^{p_{r}=0}=M(1-e/2)\neq -M/2$ and $\Omega =M(2-e)\neq -M$. The use of
$\sqrt{g_{3}}d^{3}x$ measure in (2) instead of \textquotedblleft $4\pi
R^{2}dR$" measure gives $\Omega =M(1-e)=-1.71M$, none of which obviously
coincides with the desired value. Now we do have here a traversable wormhole
with ANEC violating mass $\Omega _{ANEC}=-M$, but how can we make it
arbitrarily small? We can let $M\rightarrow 0$ to achieve it, but that would
mean that we approach the trivial Minkowski spacetime!

Let us consider a second, but qualitatively different example
provided by another class of exact solutions of the set (3) and
(4):
\[
\phi (r)=\beta \ln \left[ {\frac{{1-\frac{m}{{2r}}}}{{1+\frac{m}{{2r}}}}}%
\right] ,\quad \psi (r)=(\beta -1)\ln \left( {1-\frac{m}{{2r}}}\right)
-(\beta +1)\ln \left( {1+\frac{m}{{2r}}}\right) ,
\]%
$$
\Phi (r)=1+\left[ {\beta ^{2}-1}\right] ^{\frac{1}{2}}\ln \left[ {\frac{{1-%
\frac{m}{{2r}}}}{{1+\frac{m}{{2r}}}}}\right] .\eqno(9)
$$%
It was proposed in that form by Buchdahl [10] long ago, but it can also be
obtained from the Brans-Dicke solution by conformal rescaling. The two
undetermined constants $m$ and $\beta $ are related to the source strengths
of the gravitational and scalar parts of the configuration. The tensor mass
responsible for known gravitational effects appears always as a product $%
M_{T}=m\beta $ so that weak field effects can not separately measure the
components. Once the scalar component is set to a constant value ($\Phi
=1\Rightarrow \beta =1$), the solutions (9) reduce to the Schwarzschild
black hole in accordance with Wheeler's \textquotedblleft no scalar hair"
conjecture. Physically, this indicates the possibility that the scalar field
could be radiated away during collapse and the end result is a Schwarzschild
black hole. But for $\beta \neq 1$, the solution has a naked singularity at $%
r_{NS}=m/2$. However, the throat occurs at $r_{0}^{+}=\frac{m}{2}\left[ {%
\beta +\left( {\beta ^{2}-1}\right) ^{1/2}}\right] >r_{NS}$ and it is known
that the solution represents a traversable wormhole as it also shows $\rho
=-h,p_{r}=-h,p_{\theta }=p_{\varphi }=h$ with corresponding expression for $%
h>0$ [11]. The scalar field expands like: $\Phi \approx 1-(m/r)\sqrt{\beta
^{2}-1}+O(1/r^{2})$ and provides a charge $M_{S}=-(m/2)\sqrt{\beta ^{2}-1}$.
Using (7), we find,
$$
\Omega _{ANEC}^{p_{r}=0}=-\left( {\frac{m}{4}}\right) \times \left( {\beta
^{2}-1}\right) \times \ln \left[ {\frac{{1+1/\beta }}{{1-1/\beta }}}\right]
\approx -\left( {\frac{m}{2}}\right) \sqrt{\beta ^{2}-1}\times \left( {1-%
\frac{1}{{2\beta ^{2}}}}\right) ,\eqno(10)
$$%
from which one can read off the scalar charge. Also, like the first example,
$\Omega _{ANEC}=2\Omega _{ANEC}^{p_{r}=0}$. An interesting corollary from
Eq.(10) is the following: Consider the total energy denoted by, say, $%
\overline{M}=M_{T}+\Omega _{ANEC}^{p_{r}=0}$. At $\beta =1$, of course,  $%
\overline{M}=m$, but it turns out that, as $\beta $ increases from the value
$1$, the quantity  $\overline{M}$ decreases to a minimum value  $\overline{M}%
\approx 0.93m$ at $\beta \approx 1.16$, and again {\it increases} to  $%
\overline{M}=m$ at around $\beta \approx 1.51$. Thereafter,  $\overline{M}$
continues to grow beyond the value $m$ almost linearly with increasing $%
\beta $. These informations allow us to visualize how the total mass changes
as one increases the component of ghost energy in the configuration.
However, returning to our topic, the metric in (9) can be transformed to MTY
form under $R=re^{-\psi }$ and we can compute (2) with the \textquotedblleft
$4\pi R^{2}dR$" or any other measure but that would not give us (10). Now,
one can make $\Omega _{ANEC}\rightarrow 0$ by tuning $\beta \rightarrow 1+$,
in which case, the solution gradually approximates to the vacuum
Schwarzschild solution.

Finally, keeping in mind that the volume integral in (7) is neatly supported
by our known wormhole examples, it is curious to see what result it gives
for the \textquotedblleft $R=0$" self-dual wormhole [2,12] for which $\rho
=0 $. It is helpful to have the solution in view:

$$
ds^{2}=-\left[ \varepsilon +\lambda \left( \frac{1-\frac{m}{2r}}{1+\frac{m}{%
2r}}\right) ^{2}\right] dt^{2}+\left( 1+\frac{m}{2r}\right) ^{4}\left[ {%
dr^{2}+r^{2}d\theta ^{2}+r^{2}\sin ^{2}\theta d\varphi ^{2}}\right] ,\eqno%
(11)
$$%
where $\varepsilon $ and $\lambda $ are arbitrary constants. The
Schwarzschild solution is recovered at the value $\varepsilon =0$. The
Eq.(7) works out simply to

$$
\Omega _{ANEC}^{\rho =0}=-2m\varepsilon \ln r]_{m/2}^{\infty }\eqno(12)
$$%
If aesthetics is any guiding principle, Eq.(12) amply satisfies it when
contrasted with the expressions one obtains otherwise. It shows that it is $%
\varepsilon $ that controls the amount of ANEC violating matter.
Unfortunately, Eq.(12) together with similar expressions computed from the
metric (11) show an asymptotic logarithmic divergence.\ What does it tell
us? One possibility immediately suggests itself: Set $\varepsilon $
identically to zero, that is, conclude that an asymptotically flat spacetime
with $\rho =0$ can only be a Schwarzschild vacuum ($\Omega _{ANEC}^{\rho
=0}=0$). A more interesting possibility is to truncate the spacetime such
that the exotic matter lies only within the fixed radii $(\frac{m}{2},a]$
beyond which the spacetime is exactly Schwarzschild [2]. With Eq.(12), the
limiting arguments appear simpler and transparent. We have

$$
\Omega _{ANEC}^{\rho =0}=-2m\varepsilon \ln \left[ \frac{2a}{m}\right] ,\eqno%
(13)
$$%
so that $\Omega _{ANEC}^{\rho =0}$ $\rightarrow 0$ as $a\rightarrow m/2$
and/or $\varepsilon \rightarrow 0.$

To summarize, our key suggestion is the volume quantifier given in
Eq.(6): In the simplest case of spherical symmetry, it has
justified itself by retrieving the exact quantity of scalar charge
in the first example. Its use in the second example has thrown up
an expression for the exotic mass, viz., Eq.(10), which is not
obvious {\it a priori} and it also provides some new insights into
the behavior of total mass. When applied to the self-dual case,
Eq.(6) yields a very sensible result. Finally, as a corollary, it
is found to be consistent with the principle that the ANEC
violating matter can be made arbitrarily small [13]. It would be
worthwhile to examine the integral (6) in non-spherically
symmetric cases. This is a task for the future.\\

\noindent {\large \bf Acknowledgments} We thank Professor Matt
Visser for several enlightening comments on an earlier version of
the manuscript and Professor Rong-Gen Cai for many useful
discussions. KKN wishes to thank Professor Ou-Yang Zhong-Can for
providing hospitality and excellent working facilities at ITP,
CAS. Administrative assistance from Sun Liqun is gratefully
acknowledged. This work is supported in part by the TWAS-UNESCO
program of ICTP, Italy and the Chinese Academy of Sciences. This
work is also supported in part by National Basic Research Program
of China under Grant No. 2003CB716300 and by NNSFC under Grant No.
10175070.

\end{document}